%% file: main.tex
\documentclass[aps,prl,twocolumn,superscriptaddress,floatfix,nofootinbib,10pt]{revtex4-2}

\usepackage{amsmath,amssymb,amsfonts,stmaryrd,wasysym}
\usepackage{bm,mathrsfs,nicefrac,xfrac,braket}

\usepackage{graphicx,subfigure,float,multirow,booktabs,enumitem}
\usepackage{textcomp,url}
\usepackage{xcolor}
\usepackage{tikz}
\usepackage{quantikz}

\usepackage{bibunits}
\defaultbibliographystyle{apsrev4-2}
\defaultbibliography{references}

\usepackage{hyperref}
\hypersetup{
  colorlinks=true,
  citecolor=blue,
  urlcolor=blue,
  linkcolor=blue,
  pdfborder={0 0 0}
}

\def\bea{\begin{eqnarray}}
\def\eea{\end{eqnarray}}

\begin{document}
\begin{bibunit}
\title{An entangling Hubbard quasiparticle collider on a digital quantum processor}

\author{Chan Bin Bark}
\email{bark.13@osu.edu}
\affiliation{Department of Physics, Hanyang University, Seoul 04763, Republic of Korea}
\affiliation{Department of Physics, The Ohio State University, Columbus OH 43210, USA}

\author{Hanbyul Kim}
\affiliation{Department of Physics, Hanyang University, Seoul 04763, Republic of Korea}

\author{Yoonil Koo}
\affiliation{Department of Physics, Hanyang University, Seoul 04763, Republic of Korea}

\author{SeungBeom Hong}
\affiliation{Department of Physics, Hanyang University, Seoul 04763, Republic of Korea}

\author{Seik Pak}
\affiliation{Department of Physics, Hanyang University, Seoul 04763, Republic of Korea}

\author{Kwang Jun Ahn}
\affiliation{Department of Physics, Hanyang University, Seoul 04763, Republic of Korea}
\affiliation{Department of Energy Systems
Research, Ajou University,
Suwon 16499, Republic of Korea}

\author{Sangmo Cheon}
\email{sangmocheon@hanyang.ac.kr}
\affiliation{Department of Physics, Hanyang University, Seoul 04763, Republic of Korea}
\affiliation{Hanyang Institute for Quantum Science and Quantum Technology, Hanyang University, Seoul 04763, Republic of Korea}

\author{Youngseok Kim}
\email{kim.10017@osu.edu}
\affiliation{Department of Electrical and Computer Engineering, The Ohio State University, Columbus OH 43210, USA}

\author{Moon Jip Park}
\email{moonjippark@hanyang.ac.kr}
\affiliation{Department of Physics, Hanyang University, Seoul 04763, Republic of Korea}
\affiliation{Hanyang Institute for Quantum Science and Quantum Technology, Hanyang University, Seoul 04763, Republic of Korea}
\affiliation{Research Institute for Natural Science and High Pressure, Hanyang University, Seoul, 04763, Republic of Korea}
\date{\today}

\begin{abstract}
Scattering of quasiparticles provides direct access to the interactions that govern strongly correlated systems. In this work, we realize the quasiparticle scattering experiment of the Fermi-Hubbard model on a digital quantum processor by colliding a doublon and a holon in the half-filled chain of 21 sites encoded on 42 qubits of the \textit{ibm\_boston} processor. For the first time, we measure the transmission probability of quasiparticles of the Hubbard model on quantum hardware, which agrees with the Bethe-ansatz scattering matrix without free parameters.
Beyond the scattering probabilities, the collision generates entanglement by superposing the transmitted and reflected components, which we infer from the particle-number fluctuation. In sharp contrast, a control collision between two doublons shows neither the interaction dependence nor the number fluctuation, as expected for indistinguishable particles. Our work demonstrates that the ability to create quasiparticles of a chosen flavor on demand and program their motion and interactions makes a digital quantum processor a new experimental toolset for probing the entanglement structure generated by quasiparticle scattering.
\end{abstract}

\maketitle
 
Scattering experiments have served as a primary tool in physics for exploring the internal structure of matter. Recent advances in quantum simulators have opened the possibility of obtaining scattering data from collisions performed on quantum hardware~\cite{PRXQuantum.5.040310,halimeh_cold-atom_2025}. For instance, in lattice gauge theories, particle production and the scattering of mesons and hadrons have been simulated on trapped-ion and superconducting processors~\cite{chai_resource-efficient_2026,zemlevskiy2025scalable,farrell2026digital, halimeh2025quantumsimulationoutofequilibriumdynamics}. The fundamental advantage of quantum simulators is the tunability of the interaction strength, from weak coupling up to the strong-coupling regime where perturbative methods fail.

Strongly correlated materials are attractive systems to explore with quantum simulators.
The collisions between quasiparticles govern transport and relaxation~\cite{PhysRevLett.104.080401,PhysRevB.82.224302,RevModPhys.83.471,RevModPhys.70.1039}, yet they are accessed only indirectly through spectroscopy~\cite{PhysRevLett.120.166401,Giannetti_2016}. A quantum simulator can instead prepare individual quasiparticles and resolve their collision both spatially and temporally. For instance, entanglement production through scattering, which cross sections do not capture, can be probed directly~\cite{PhysRevD.104.114501,Entanglement_nature,doi:10.1126/science.aau4963}. Recent advances in error mitigation further extend such time-resolved simulations to increasingly long times on current quantum hardware~\cite{hartnett2026fastaccuratehighresolutionsimulation,hayata_digital_2026,shinjo2026unveiling,chen2026robust}.

\begin{figure*}[t]
\centering
\includegraphics[width=\textwidth]{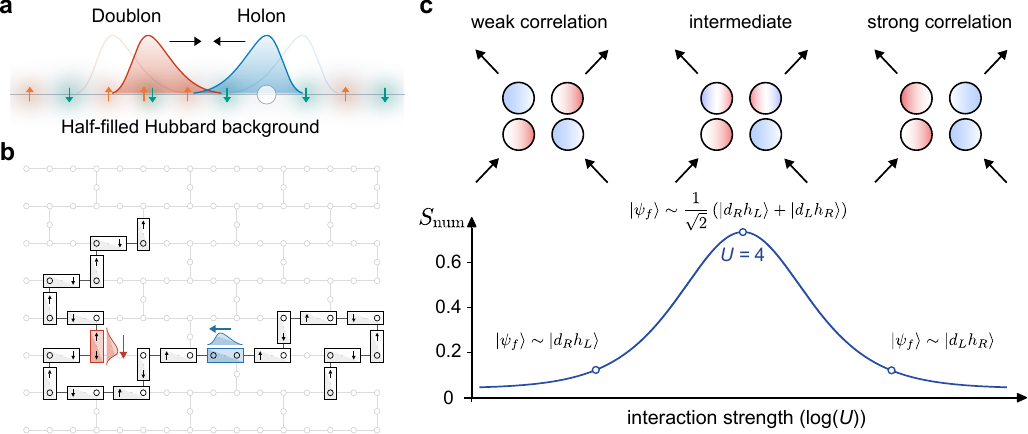}
\caption{\textbf{Doublon--holon collider on a 42-qubit chain.} (a) Conceptual figure of the doublon--holon collision. (b) Qubit layout on the \textit{ibm\_boston} processor and the initial state. Two neighboring qubits in the chain (box) form one Hubbard site, and the two qubits correspond to the spin-up and spin-down orbitals, respectively. Black arrows denote the spin direction of an occupied orbital, and empty circles denote empty orbitals. On the half-filled N\'eel background, a doublon (red) is prepared on one side and a holon (blue) on the other side. (c) Schematic of the entanglement structure generated by quasiparticle scattering. Top row: Depending on the interaction strength, the two quasiparticles either transmit through each other (weak), reflect off each other (strong), or form an entangled superposition of transmitted and reflected states (intermediate). Bottom row: Number entropy $S_\text{num}$, defined in equation~\eqref{eq:number_entropy}, as a function of the interaction strength $U$ (logarithmic scale). The entanglement and $S_\text{num}$ reach their maximum at $U/J = 4$.}
\label{fig:schematic}
\end{figure*}

Although the quench dynamics of the Fermi–Hubbard model have been simulated on quantum processors~\cite{arute2020observation,alam2025programmable,hartnett2026fastaccuratehighresolutionsimulation}, collisions between individually prepared distinct quasiparticles have not yet been realized. Such collisions differ qualitatively from those between identical (quasi)particles, such as doublons or magnons~\cite{andersen2026precisionquantumsimulationmagnon,wang2025observationinelasticmesonscattering,farrell2026digital,halimeh_cold-atom_2025}: the collision dynamics of a doublon and a holon are intrinsically governed by correlation strengths, and may therefore offer deeper insight into strongly correlated phenomena. In this work, we realize a quasiparticle collider with a digital quantum processor. We collide a doublon and a holon, the charge excitations of the half-filled Mott insulator, on a chain of 21 sites encoded on 42 qubits of the superconducting quantum processor \textit{ibm\_boston} (Fig.~\ref{fig:schematic}a,b). As the interaction strength increases, the collision changes from transmission-dominated to reflection-dominated scattering. The two-body scattering matrix of the one-dimensional Fermi-Hubbard model is known exactly from the Bethe ansatz~\cite{PhysRevLett.20.1445, PhysRevLett.56.2453, PhysRevLett.72.908,7ef6589804f8480e8e84711e5ad18a16}, which provides a calibration for our collider. For the first time, we measure the transmission probability of Hubbard quasiparticles on quantum hardware, and it agrees with the Bethe-ansatz scattering matrix from weak to strong coupling without free parameters. Furthermore, the collision superposes the transmitted and reflected components, which entangles the left and right parts of the chain (Fig.~\ref{fig:schematic}c). We infer this entanglement from the particle-number fluctuation without state tomography. We emphasize that both the interaction dependence and the number fluctuation are essentially absent in a control collision between two doublons, as expected for indistinguishable particles. These signatures are therefore specific to the doublon–holon pair, a particle–antiparticle pair on the Mott background whose exchange channel is controlled by the interaction strength, and isolating them requires freely choosing the colliding flavors and tuning the interaction, a degree of control uniquely available in a digital quantum simulation.

\section{Results}
\subsection{A doublon–holon collider on a digital quantum processor} Figure~\ref{fig:schematic}a exemplifies our collider setup, in which a doublon and a holon propagate toward each other from the two sides of the chain. The doublon and the holon are the charge excitations on the Mott background of the half-filled Fermi--Hubbard chain,
\begin{align}
    H = &-J \sum_{i=1}^{L-1} \sum_{\sigma}
        \left( c_{i,\sigma}^{\dagger} c_{i+1,\sigma} + \mathrm{h.c.} \right) \nonumber \\
        &+ U \sum_{i=1}^{L} \left(n_{i,\uparrow}-\tfrac{1}{2}\right)\left(n_{i,\downarrow}-\tfrac{1}{2}\right),
    \label{eq:fermi_hubbard}
\end{align}
where $c_{i,\sigma}^{\dagger}$ ($c_{i,\sigma}$) creates (annihilates) a fermion with spin $\sigma \in \{\uparrow,\downarrow\}$ on site $i$, and $n_{i,\sigma} = c_{i,\sigma}^{\dagger} c_{i,\sigma}$ is the corresponding number operator. Here, $J$ is the nearest-neighbor hopping amplitude and $U$ is the on-site interaction strength. Throughout this work, we set $\hbar = J = 1$ for convenience, so that energies are measured in units of $J$ and times $T$ in units of $\hbar/J$. Starting from a N\'eel state($\ket{\uparrow \downarrow \uparrow \cdots \downarrow \uparrow } $), we locally create a doublon and a holon on two sites equidistant from the center of the chain (See Methods for the site convention).

To launch the two quasiparticles toward each other with finite momenta, we divide the time evolution into two phases (see Extended Data Fig.~\ref{fig:setup}). In the first phase ($0 < T \le 1$), the hopping from each excitation site to the nearer chain edge is deactivated. Each quasiparticle propagates only toward the center of the chain as a directed wave packet (Fig.~\ref{fig:main_result}a--c). In the second phase ($1 < T \le 4$), all hopping terms are restored, and the two wave packets move toward each other and collide. When the doublon and the holon are treated as free particles with hopping $J$, the wave packet formed in phase~1 is given by the exact solution of a semi-infinite chain in terms of Bessel functions~\cite{Hartmann_2004}. By comparing with the Bessel function solution, we find that more than 93\% of its weight is carried by the leading packet for $T \le 4$. The trailing packets are negligible on our time scale. The detailed wave-packet profiles are given in Supplementary Material.

We implement this chain with $L=21$ sites on 42 qubits of the IBM superconducting quantum processor \textit{ibm\_boston} (Fig.~\ref{fig:schematic}b). Fermionic operators are mapped onto qubits by the Jordan--Wigner transformation~\cite{JWtransform} with a zigzag ordering~\cite{hartnett2026fastaccuratehighresolutionsimulation}, in which same-spin orbitals of adjacent sites occupy physically neighboring qubits. With this ordering, each Trotter step requires only nearest-neighbor gates (see Methods and Supplementary Material for the detailed implementation). To reduce the Trotter error within the limited circuit depth, we fix the number of Trotter steps to $N=20$ and adaptively refine the time steps around the collision. The adaptive time mesh substantially lowers the Trotter error at late times (Supplementary Material, Fig.~\ref{figS4}). The hardware error is suppressed by employing Pauli twirling and dynamical decoupling. The remaining errors are mitigated in post-processing using the particle-number conservation and the inversion--particle-hole symmetry of the problem (see Methods). Hereafter, we refer to this post-processing as \textit{mitigation}.

Since the initial N\'eel state is not the exact ground state of the Hamiltonian, the measured values can be affected by various effects unrelated to the quasiparticle dynamics such as edge effects or offsets that depend on the device error rates. To remove these contributions, for each measurement we performed a reference run using the same circuit without initial excitation and subtracted its measured value. In the following, such background-subtracted quantities are denoted with a prefix $\Delta$, $\Delta A = A - A_{\rm bg}$, where $A_{\rm bg}$ denotes the same quantity $A$ obtained without the doublon and holon excitations.
\subsection{Experimental results and S-matrix reconstruction} 

\begin{figure*}[t]
\centering
\includegraphics[width=\textwidth]{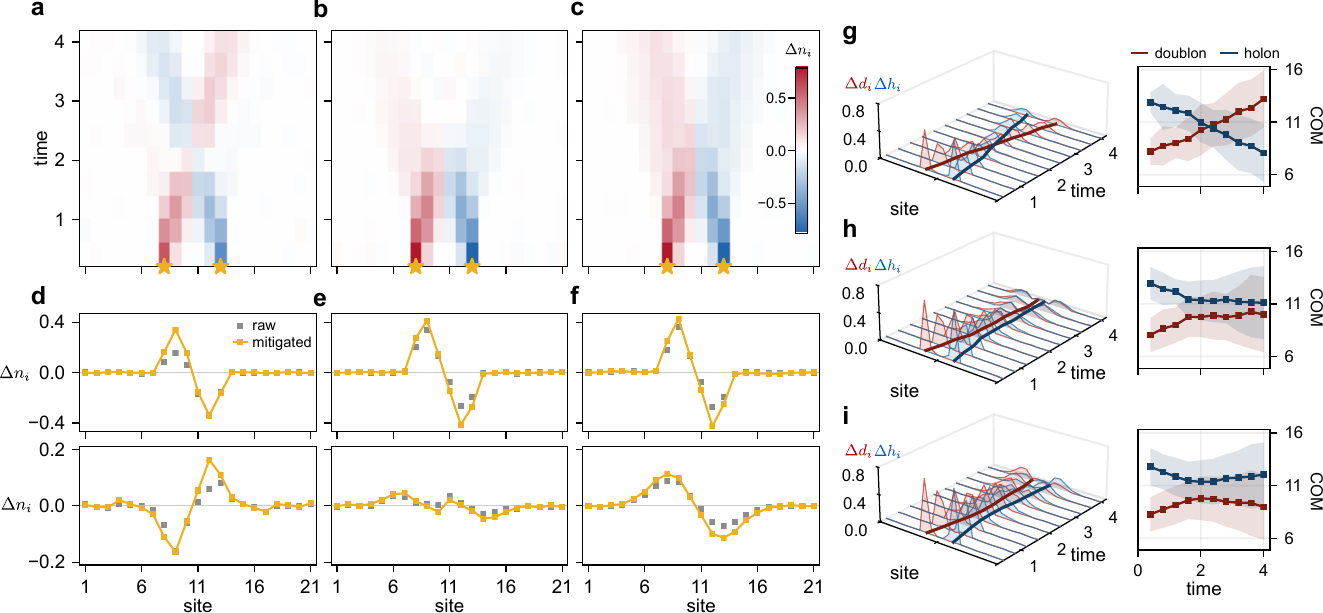}
\caption{\textbf{Charge dynamics of the doublon--holon collision.} (a--c) Charge density $\Delta n_{i}(T)$ of the doublon--holon collision as a function of time for interaction strength $U=1$ (a), $3.5$ (b), and $7$ (c). The evolution time is shown for 10 time points from $T=0.4$ to $4.0$, and the initial positions of the doublon and holon excitations are marked by stars. (d--f) Site-resolved charge density for $U=1$ (d), $3.5$ (e), and $7$ (f). The top row shows the experimental results at $T=1.2$ (before the collision) and the bottom row at $T=3.2$ (after the collision); gray plots are the raw data and yellow plots are the results after mitigation. (g--i) For $U=1$ (g), $3.5$ (h), and $7$ (i), time-resolved profiles of the site-resolved doublon and holon number $\Delta d_{i}$ (red) and $\Delta h_{i}$ (blue) (left), and the center of mass (line) and the packet width (standard deviation, shaded region) of each as a function of time (right).}
\label{fig:main_result}
\end{figure*}

Figures~\ref{fig:main_result}a--f show the space-time maps of the charge density measured for $U=1$, $3.5$, and $7$, as schematically illustrated in a top row of Fig.~\ref{fig:schematic}c. We first observe well-defined wave packets of the doublon (red) and the holon (blue) propagating toward each other and colliding at the center of the chain around $T \approx 2$. For all values of $U$, there is no noticeable change in the speed of the outgoing packets relative to the incoming ones. This indicates that the collision is elastic, which is consistent with the integrability of the one-dimensional Hubbard model~\cite{PhysRevLett.20.1445,PhysRevLett.56.2453,7ef6589804f8480e8e84711e5ad18a16}. As $U$ increases, we find that the collision changes from transmission-dominated to reflection-dominated scattering. At $U=1$, the doublon and holon packets pass through each other and maintain their original directions (Fig.~\ref{fig:main_result}a,d). Contrarily, at $U=7$, the two packets are reflected back upon collision (Fig.~\ref{fig:main_result}c,f). At the intermediate value $U=3.5$, the charge density nearly vanishes after the collision (Fig.~\ref{fig:main_result}b,e), which we discuss in detail below. The crossover between the two regimes is already visible in the raw data, and the mitigation mainly restores the amplitude (Fig.~\ref{fig:main_result}d--f).

This crossover can be compared with the two-body scattering matrix of the one-dimensional Hubbard model, which is obtained from the Bethe ansatz. For the doublon--holon pair, the transmission probability is
\begin{align}
\mathcal{T}(\theta) = \frac{\theta^{2}}{1+\theta^{2}},
\label{transmission}
\end{align}
with $\theta = 2J\left(\sin k_{1} - \sin k_{2}\right)/U$, where $k_{1}$ and $k_{2}$ are the momenta of the two colliding wave packets. The numerator of $\theta$ is the relative group velocity $v_{\mathrm{rel}}$ of the two packets, since the group velocity of a single quasiparticle is $v(k) = 2J\sin k$ (See the Supplementary Material for derivation.) The observed crossover is consistent with equation~\eqref{transmission}. For weak interaction ($U\sim0$), $\mathcal{T}$ approaches unity. Contrarily, for strong interaction, $\mathcal{T} \simeq \theta^{2} = (v_{\mathrm{rel}}/U)^{2}$ decreases rapidly with $U$, and the collision becomes fully reflective. 

This behavior can be understood from strong-coupling perturbation theory. For the doublon (${\uparrow\!\downarrow}$) and the holon ($\circ$) to exchange their positions, the pair must pass through an intermediate configuration with the recombinations,
$
|\cdots{\uparrow},{\uparrow\!\downarrow},{\circ},{\downarrow}\cdots\rangle
\;\to\;
|\cdots{\uparrow},{\uparrow},{\downarrow},{\downarrow}\cdots\rangle
\;\to\;
|\cdots{\uparrow},{\circ},{\uparrow\!\downarrow},{\downarrow}\cdots\rangle,
$
This intermediate configuration is detuned by $U$ from the initial and final states, and the exchange amplitude is therefore suppressed to an order of $J^{2}/U$. As a result, the two quasiparticles reflect off each other as hard-core particles.

\begin{figure}[t]
\centering
\includegraphics[width=0.48\textwidth]{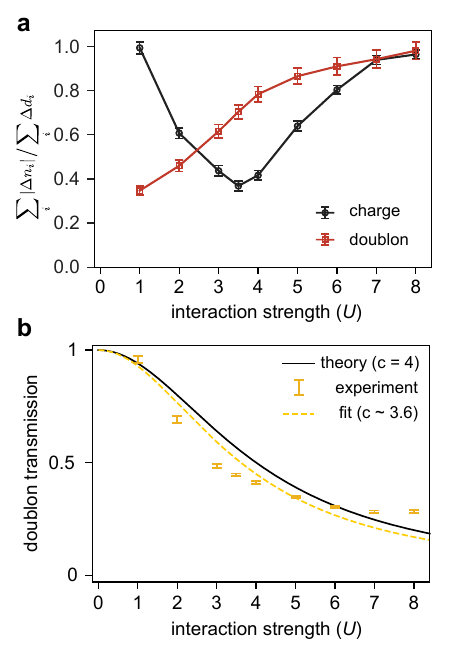}
\caption{\textbf{Interaction dependence of the collision dynamics.} Experimental results (mitigated) at $T=3.2$ for $U=1,2,\cdots,8$ and $U=3.5$. (a) Total charge $\sum_{i}|\Delta n_{i}|$ (black) and total doublon number $N_{d}=\sum_{i}\Delta d_{i}$ (red). (b) Doublon transmission probability as a function of the interaction strength. The transmission probability $\mathcal{T}$ is defined as the fraction of the total doublon number $N_{d}$ located on the right side ($i\geq 11$) of the collision center. Yellow error bars are the experimental values, the black solid line is the theoretical curve with $c=4$, and the yellow dashed line is the fit of equation~(\ref{transmission}) to the experimental values ($c\sim 3.6$).}
\label{fig:transmission}
\end{figure}

The whole crossover can be made quantitative by extracting the transmission probability directly from the doublon density $\Delta d_i=\Delta \langle n_{i\uparrow}n_{i\downarrow}\rangle$. After the collision ($T=3.2$), we extract $\mathcal{T}$ as the fraction of the total doublon number found on the far side of the chain center (Fig.~\ref{fig:transmission}b). Substituting the maximum relative group velocity $c = v_{\mathrm{rel}} = 4J$ into equation~\eqref{transmission} leads to $\mathcal{T} = (c/U)^{2} / [1 + (c/U)^{2}]$ and the $U$-dependent behavior of the experimental data agrees well with with the theoretical curve (black solid line) without free parameters. Fitting the experimental data to equation~\eqref{transmission} with $c$ as a free parameter gives $c \approx 3.6J$, an effective velocity smaller than the maximum relative velocity. This is because the packet formed in phase~1 contains a spread of momenta around $k=\pi/2$, so that their average relative velocity is smaller than $4J$.

\subsection{Superposition of transmission and reflection channel} 
Focusing on the intermediate regime ($U=3.5$, Fig.~\ref{fig:main_result}b), we find that the charge density of the two packets nearly vanishes after the collision. The vanishing density indicates that the superposition of the transmitted and the reflected components with comparable weights,
\begin{equation}
|\psi_{\rm out}\rangle \sim t\,|d_{R}\,h_{L}\rangle + r\,|d_{L}\,h_{R}\rangle ,
\label{eq:superposition}
\end{equation}
where $|d_{R}h_{L}\rangle$ denotes the doublon packet on the right and the holon packet on the left of the collision center, and $|d_{L}h_{R}\rangle$ denotes the reverse. Here, $|t|^{2}=\mathcal{T}$ and $|r|^{2}=1-\mathcal{T}$. At $\mathcal{T}\approx1/2$, each side of the chain is described by equal superpositions of the doublon and the holon. Their opposite charges then cancel on each side, and the charge density vanishes. The same vanishing, however, would also arise if the pair annihilation of the doublon--holon pair occurs, which we examine the possbility below.

To distinguish the two possibilities, we separately measure the doublon density $d_i=\langle n_{i\uparrow}n_{i\downarrow}\rangle$ and the holon density $h_i=\langle(1-n_{i\uparrow})(1-n_{i\downarrow})\rangle$. Figures~\ref{fig:main_result}g--i show $\Delta d_i$ and $\Delta h_i$ together with the center of mass and the width of each distribution. At $U=1$, the centers of mass of the doublon and the holon cross each other, whereas at $U=7$ they turn back toward their initial sides. At the intermediate value $U=3.5$, both centers of mass remain near the center of the chain after the collision, while the widths of the two distributions broaden. This indicates that the doublon and the holon each spread over both sides of the chain, as described by equation~\eqref{eq:superposition}.

To make this quantitative, Fig.~\ref{fig:transmission}a compares the total charge, $\sum_i |\Delta \langle n_i \rangle|$, with the total doublon number, $\sum_i \Delta d_i$, after the collision ($T = 3.2$). The total charge shows a dip at $U \approx 3.5$, whereas the doublon number increases monotonically with $U$. Recombination of the doublon--holon pair would reduce the doublon number together with the charge. The absence of non-monotonicity in the total doublon number therefore rules out recombination. The monotonic increase of the doublon number itself reflects the decay of the created doublon into the background at small $U$, which is independent of the collision and is suppressed as the Mott gap opens. Since the transmission probability is normalized by the total doublon number, this decay does not affect the comparison with equation~\eqref{transmission}.

\subsection{Number entanglement generated by the collision}

The superposed state in equation~\eqref{eq:superposition} generates the entanglement between the left and right part of the chain. Since the particle number is conserved, the entanglement entropy between the two halves resolves into a number entropy and a configurational entropy, $S = S_{\rm num} + S_{\rm conf}$. The number entropy $S_{\rm num}$ arises from the fluctuation of the particle number $N_L$ in the left half of the chain($i\leq10$) while the configurational entropy $S_{\rm conf} $ arises from superpositions within each particle-number sector~\cite{PhysRevLett.102.100502,PhysRevB.82.012405,PhysRevB.85.035409,doi:10.1126/science.aau0818}. The transmitted and reflected components in equation~\eqref{eq:superposition} differ in $N_L$ by two, and the collision generates the number entropy
\begin{equation}
S_{\rm num}=-\mathcal{T}\ln\mathcal{T}-(1-\mathcal{T})\ln(1-\mathcal{T}),
\label{eq:number_entropy}
\end{equation}
which reaches $\ln 2$ at $\mathcal{T}=1/2$. The dependence of $S_{\rm num}$ on the interaction strength is shown in Fig.~\ref{fig:schematic}c. The variance of the particle number, $\mathrm{Var}(N_L)=4\mathcal{T}(1-\mathcal{T})$, which monotonically corresponds to $S_{\rm num}$ is also maximal at $\mathcal{T}=1/2$ and is obtained directly from the $Z$-basis measurements~\cite{PhysRevLett.102.100502, PhysRevB.82.012405,PhysRevB.85.035409}. A detailed discussion can be found in our supplementary information.

\begin{figure}[t]
\centering
\includegraphics[width=0.48\textwidth]{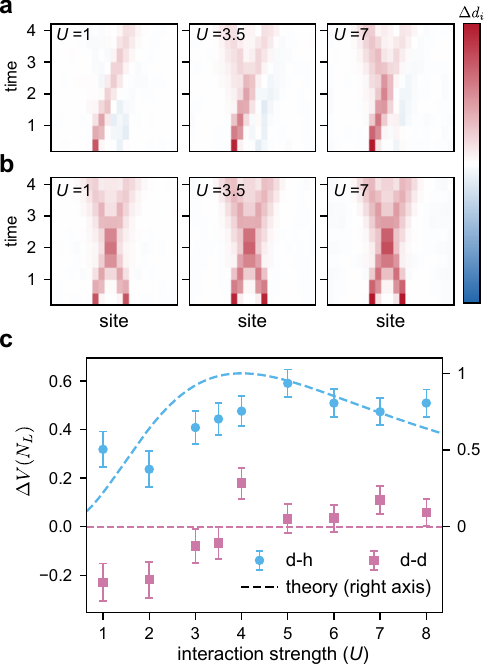}
\caption{\textbf{Comparison of the doublon--holon(d-h) and doublon--doublon(d-d) collisions.} (a, b) Heat maps of the site-resolved doublon density $\Delta d_{i}$ as a function of time for various interaction strengths in the d-h collision (a) and the d-d collision (b). The evolution time is shown for 10 time points from $T=0.4$ to $4.0$. (c) Interaction dependence of the particle-number variance of the left-half sites, $\Delta V(N_{L})$, at the reference time $T=3.2$ after the collision. Blue points are for the d-h collision and pink points for the d-d collision. The dashed lines show the $\mathrm{Var}(N_L) = 4\mathcal{T}(1-\mathcal{T})$ for d--h and $0$ for d--d, plotted on the right axis.}
\label{fig:doublon_doublon}
\end{figure}

For comparison, we replace the holon by a second doublon and repeat the collision under the same conditions. For two identical doublons, the exchange channel is absent because of the Pauli exclusion principle. Indeed, the doublon--doublon collision shows nearly identical dynamics for all $U$, apart from a small $U$-dependent decay of the doublons (Fig.~\ref{fig:doublon_doublon}a,b). The $U$ dependence of the doublon--holon collision therefore originates from the exchange channel of the particle--antiparticle pair on the Mott background.

Figure~\ref{fig:doublon_doublon}c compares the variance of the particle number in the left half after the collision ($T = 3.2$) for the two cases. For the doublon--doublon collision, the variance stays close to zero for all $U$. Contrarily, the doublon--holon collision gives a large variance, which indicates that the collision splits the state into two components with different particle numbers in the left half. We emphasize that this collision-generated number entanglement is measured without state tomography, using only the site-resolved occupations. This measurement, however, shows only the presence of the two components and their relative weights, and does not directly confirm the coherence between them.

\section{Discussion} 

Our results demonstrate that digital quantum processors can serve as a promising experimental platform for scattering experiments designed at the level of individual quasiparticles. As a first time, our hardware experiment reproduces the scattering matrix of one-dimensional Hubbard model predicted by Bethe ansatz. This measurement infer the possible long-range superposed states described by the transmission and reflection channel~\cite{bao2015universal}. However, the measured data shows only the presence of the two components and their relative weights, and does not directly confirm the coherence between them. The interpretation as a coherent superposition is based on the fact that the time evolution is unitary and that the observed $U$ dependence follows the prediction of equation~\eqref{transmission}, and it could be confirmed by an additional interference experiment on the wave function after the collision once longer simulation times become accessible.

\newpage
\pagebreak

\section{Methods}

\textbf{Wave-packet preparation.} Since a locally created excitation spreads ballistically in both directions under time evolution (Extended Data Fig.~\ref{fig:setup}a,b), realizing a well-defined collision requires wave packets with finite momenta moving toward each other. To this end, we divide the time evolution into two phases. In the first phase (phase~1), the hopping toward the boundary is deactivated at each excitation site while the system evolves for a time $T = 1$, so that the two excitations form directed packets propagating only toward the center of the chain (Extended Data Fig.~\ref{fig:setup}c,d). The wave function of the packet formed in this way is described by Bessel functions of the first kind; within the time window of interest in this work, the leading packet component is sufficiently dominant that it behaves as a single well-defined wave packet (see Supplementary Material). In the second phase (phase~2), all hopping terms are activated and the system evolves for the remaining time, so that the two quasiparticles collide as packets carrying finite momenta. The total simulation time is $T = 4$. The simplified circuit structure of each phase is shown in Extended Data Fig.~\ref{fig:setup}e. A single doublon (or holon) propagates ballistically on this background like a free particle with an effective hopping amplitude $J$, and the propagated wave function can also be described by Bessel functions of the first kind.

\textbf{Error suppression.} To suppress the hardware error, we first select the qubit layout carefully. Qubits and couplers whose readout and CZ error rates exceed threshold values in the calibration data are excluded as a hard constraint, and Bell-state fidelity tests on all remaining couplers determine the optimal 42-qubit chain~\cite{mckay2023benchmarkingquantumprocessorperformance, cosentino2026metriqcollaborativeplatformbenchmarking}. Each data point is then measured with a total of 70,000 shots divided into 70 Pauli twirling instances~\cite{PhysRevA.94.052325}, combined with dynamical decoupling~\cite{PhysRevA.58.2733,PhysRevLett.121.220502,PhysRevApplied.20.064027} and readout twirling~\cite{PhysRevA.105.032620,doi:10.1126/sciadv.abi8009}, which approximately convert coherent gate errors and asymmetric readout errors into stochastic bit-flip channels. As the observable, we measure the $Z-$basis snapshots of each qubit and reconstruct the site-resolved charge, doublon, and holon densities through the qubit--site mapping.

\textbf{Post-processing.} Our model Hamiltonian conserves the total particle number $N_{\mathrm{tot}}$ and that the total particle number of the initial state is known, it can give useful information which is regarded on the hardware noise~\cite{PhysRevA.98.062339, Cai_2021}. We estimated the hardware error rate from the distribution of the total particle number and used it to restore the noiseless expectation values from the noisy results. In addition to the spin SU(2) symmetry, the model possesses a pseudospin SU(2) symmetry (also known as $\eta$-pairing) in the charge sector, under which the doublon and the holon are dual to each other~\cite{PhysRevLett.63.2144}. The Hamiltonian is furthermore invariant under the spatial inversion about the chain center, $\mathcal{I}: i \rightarrow L+1-i$, as well as under its combination $\mathcal{IC}$ with the particle-hole transformation $\mathcal{C}: c_{i,\sigma} \rightarrow (-1)^i c_{i,\sigma}^{\dagger}$. In our experiment, we excites a doublon on one side of the chain center and a holon on the other initially, so that inversion symmetry alone is not satisfied to entire time evolution. However, by pseudospin symmetry, invariant under the combined symmetry $\mathcal{IC}$ is satisfied and this symmetry can be exploited in the data post-processing to reduce errors from the device noises. In our experiment, the excitations and the symmetrization are defined with respect to the first 20 sites to make the length even. The \textit{center} refers to the center of this 20-site chain (between sites 10 and 11), and $\mathcal{I}: i \rightarrow 21-i$.

\textbf{Trotterization.} The time evolution is implemented with the following custom Trotter decomposition. A single Trotter step consists of a hopping layer acting on one set of spins, a layer combining the on-site interaction with fermionic swaps, and another hopping layer. Because the fermionic swaps exchange the positions of the spin-up and spin-down qubits, the second hopping layer applies hopping to the other set of spins with an identical gate arrangement. Since an odd number of steps leaves the spin-up and spin-down positions interchanged with respect to the initial layout, we always use an even number of steps. The hopping layer is implemented with a two-qubit depth of two CZ gates, and the interaction--fswap layer with three CZ gates via a KAK decomposition~\cite{KHANEJA200111,PhysRevA.69.032315,PhysRevA.69.010301}, each supplemented by single-qubit gates. As the last hopping layer of each step merges with the first hopping layer of the next, the time evolution of $N$ steps is realized with a two-qubit circuit depth of $5N + 2$. We simulate up to 20 Trotter steps. Details of the mapping and circuit construction are provided in the Supplementary Material.

\textbf{Hamiltonian parameters.} For the interaction strength, we fixed $J=1$ and scan $U=1$, $3.5$, and $7$ when tracking the full time dynamics, and $U=1$ through $8$ in integer steps when probing the $U$ dependence of the scattering outcome at a fixed time shortly after the collision ($T=3.2$).

\textbf{Numerical simulations.} We compared the experimental results with tensor-network simulations (MPS-based TEBD simulation), performed using \texttt{TeNPy}~\cite{tenpy2024}. The agreement is sufficient to clearly distinguish the different dynamical behaviors. Comparisons between the experimental and numerical data are shown in the Extended Data Figs.~\ref{fig:numerical1}--\ref{fig:numerical4}.

See supplementary materials for details of the methods used in this work.

\medskip
\noindent\textbf{Data availability}\\
All key experimental data supporting the findings of this study are provided in the Article. The raw datasets are available from the corresponding author upon request for the purposes of academic replication and review.

\medskip
\noindent\textbf{Code availability}\\
The code used for analyses is available from the corresponding author upon request.

\medskip
\noindent\textbf{Acknowledgements}\\
Y. Kim and C.B.B. acknowledge the Center for Quantum Information Science and Engineering (CQISE) at The Ohio State University for its support and resources. This work was supported by the National Research Foundation of Korea (NRF) grants funded by the Korea government (MSIT) (Nos.~RS-2025-16070482, RS-2025-25464760, RS-2026-25607155, RS-2025-03392969, RS-2025-25446099, and RS-2023-NR119928), and by the Institute of Information \& Communications Technology Planning \& Evaluation (IITP) grant funded by the Korea government (MSIT) (No.~RS-2026-25519864).

\medskip
\noindent\textbf{Author contributions}\\
C.B.B. performed the experiments on the IBM quantum processor. C.B.B. analyzed the data with input from H.K. C.B.B. performed the tensor-network calculations. H.K. prepared the conceptual figures. H.K. also contributed to the theoretical interpretation of the scattering results. Y. Koo, S.P. and S.B.H. developed the qubit mapping and the circuit implementation of the Trotter evolution. K.J.A. contributed to the analysis of the results. S.C. and Y. Kim supervised the project. Y. Kim guided the hardware experiments and the error mitigation. M.J.P. conceived the project. M.J.P. supervised the overall research. C.B.B., H.K. and M.J.P. wrote the manuscript with input from all authors.

\medskip
\noindent\textbf{Competing interests}\\
The authors declare no competing interests.

\medskip
\noindent\textbf{Correspondence}\\
Correspondence and requests for materials should be addressed to Sangmo Cheon, Youngseok Kim or Moon Jip Park.

\putbib

\end{bibunit}


\clearpage
\onecolumngrid

\begin{center}
\textbf{\large Extended Data Figures}
\end{center}
\setcounter{figure}{0}
\renewcommand{\figurename}{Extended Data Fig.}
\renewcommand{\theHfigure}{ED\arabic{figure}}

\begin{figure*}[h]
\centering
\includegraphics[width=\textwidth]{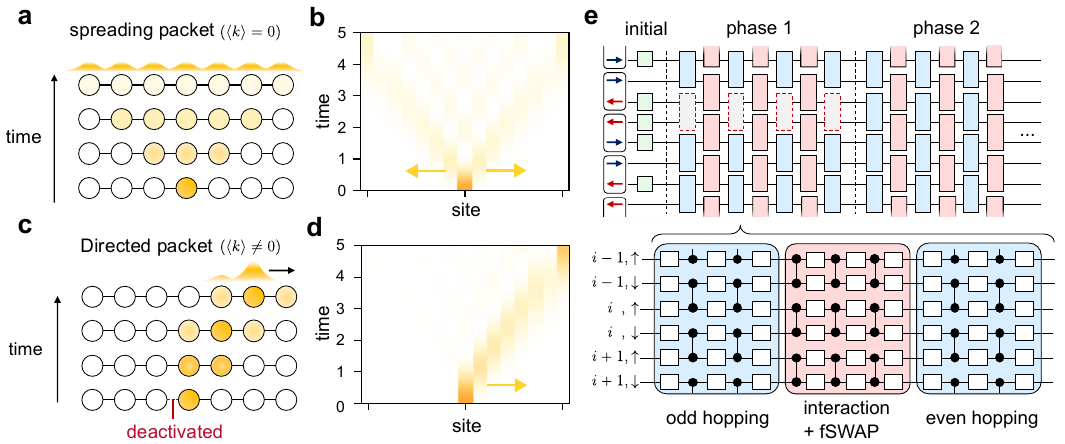}
\caption{\textbf{Experimental setup}. (a--d) Propagation of the wave function when the left hopping is initially kept (a, b) and when it is deactivated (c, d). (a, c) are schematics, and (b, d) show the site-resolved amplitude $|\psi|^{2}$ obtained from the exact time evolution of a single particle placed at the central site of an empty lattice. (e) Schematic of the electron--qubit mapping, the overall experimental process, and the quantum circuit for time evolution. Gray boxes in Phase~1 denote deactivated hoppings.}
\label{fig:setup}
\end{figure*}

\begin{figure*}[t]
\centering
\includegraphics[width=0.7\textwidth]{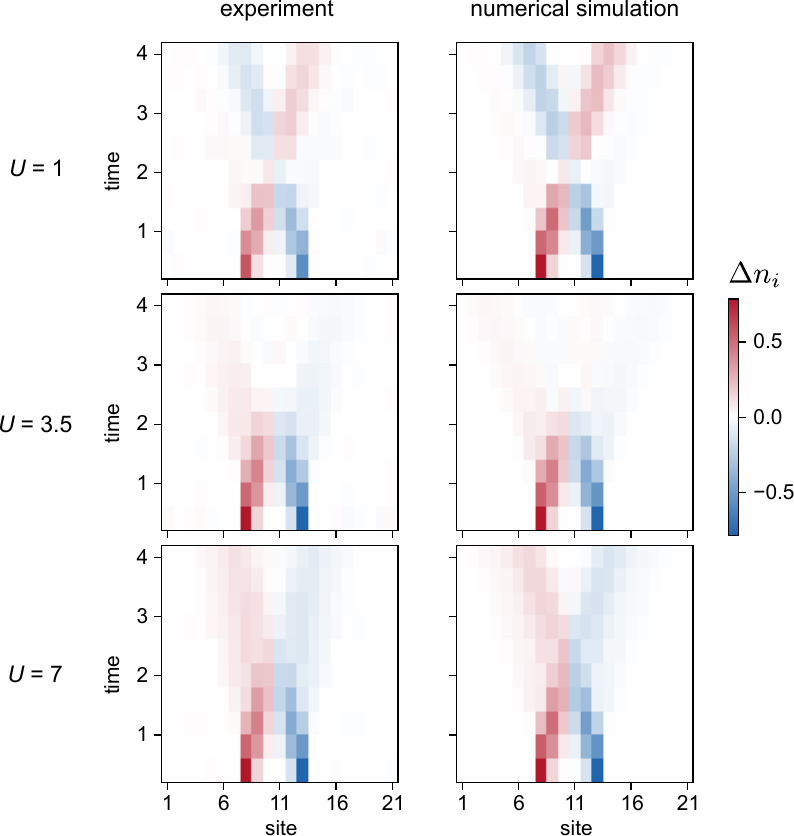}
\caption{\textbf{Comparison of experimental data and numerical simulations of the charge density dynamics in the d--h collision.} The numerical simulations (right) are performed with the same parameters as the experiment (left). This figure supplements the data in Fig.~\ref{fig:main_result}(a)--(c).}
\label{fig:numerical1}
\end{figure*}

\begin{figure*}[t]
\centering
\includegraphics[width=0.7\textwidth]{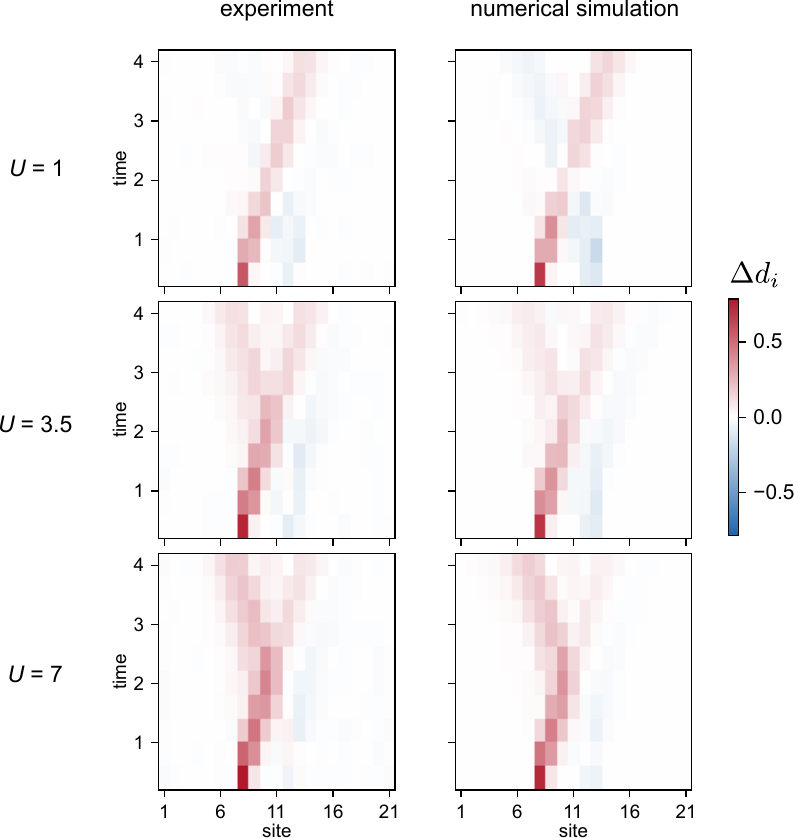}
\caption{\textbf{Comparison of experimental data and numerical simulations of the doublon density dynamics in the d--h collision.} The numerical simulations (right) are performed with the same parameters as the experiment (left). This figure supplements the data in Fig.~\ref{fig:doublon_doublon}(a).}
\label{fig:numerical2}
\end{figure*}

\begin{figure*}[t]
\centering
\includegraphics[width=0.7\textwidth]{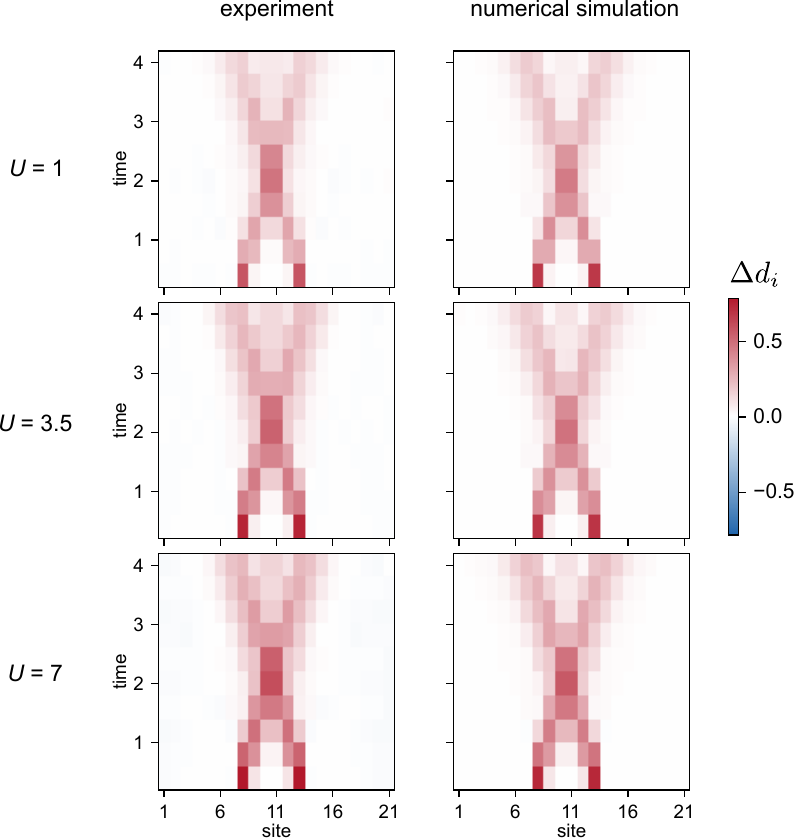}
\caption{\textbf{Comparison of experimental data and numerical simulations of the doublon density dynamics in the d--d collision.} The numerical simulations (right) are performed with the same parameters as the experiment (left). This figure supplements the data in Fig.~\ref{fig:doublon_doublon}(b).}
\label{fig:numerical3}
\end{figure*}

\begin{figure*}[t]
\centering
\includegraphics[width=\textwidth]{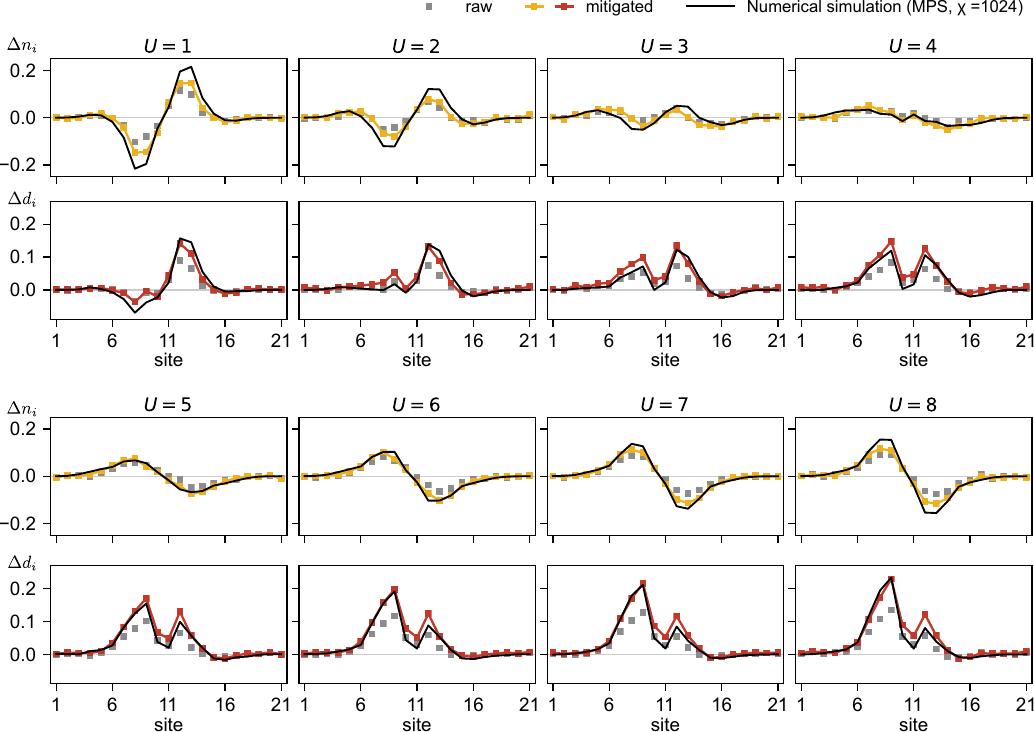}
\caption{\textbf{Comparison of experimental data and numerical simulations of the site-resolved charge and doublon densities in the d--h collision.} Site-resolved densities at $T = 3.2$ are shown for $U = 1, \dots, 8$. Gray markers show the raw charge density, and yellow and red markers show the error-mitigated charge and doublon densities, respectively. Black solid lines show the corresponding MPS simulations with bond dimension $\chi = 1024$. These plots supplement the data in Fig.~\ref{fig:main_result}(d)--(f) and Fig.~\ref{fig:transmission}.}
\label{fig:numerical4}
\end{figure*}

\clearpage
\onecolumngrid

\begin{center}
\textbf{\large Supplementary Material}
\end{center}
\renewcommand{\figurename}{FIG.}

\setcounter{equation}{0}
\setcounter{figure}{0}
\setcounter{table}{0}
\setcounter{section}{0}
\renewcommand{\theequation}{S\arabic{equation}}
\renewcommand{\thefigure}{S\arabic{figure}}
\renewcommand{\thetable}{S\arabic{table}}
\renewcommand{\thesection}{S\arabic{section}}
\renewcommand{\theHequation}{S\arabic{equation}}
\renewcommand{\theHfigure}{S\arabic{figure}}
\renewcommand{\theHtable}{S\arabic{table}}
\renewcommand{\theHsection}{S\arabic{section}}

\begin{bibunit}
\renewcommand{\citenumfont}[1]{S#1}
\renewcommand{\bibnumfmt}[1]{[S#1]}
\renewcommand{\bibsection}{\section*{Supplementary References}}

\input{supple_include}

\putbib
\end{bibunit}

\end{document}

%% file: supple_include.tex
\setcounter{section}{0}
\renewcommand{\thesection}{S\arabic{section}}
\setcounter{figure}{0}
\renewcommand{\thefigure}{S\arabic{figure}}
\setcounter{table}{0}
\renewcommand{\thetable}{S\arabic{table}}
\setcounter{equation}{0}
\renewcommand{\theequation}{S\arabic{equation}}

\setcounter{figure}{0}

\section{Wave-packet preparation in phase 1}\label{sec:phase1}
In this section, we analytically describe the wave packet prepared by deactivating the hopping terms in phase 1. Similar to the preparation technique previously used in cold-atom simulators, where propagation is restricted by a strong potential barrier~\cite{PRXQuantum.5.040310}, the short-time dynamics of phase 1 is that of a particle initially excited at the edge of a semi-infinite one-dimensional chain. For a particle initially localized at site $n_0$ of an infinite chain with hopping amplitude $J$, the wave function at site $n$ and time $t$ is written as
\begin{align}
    \psi_n(t) = i^{\,n-n_0} J_{n-n_0}(2Jt),
\end{align}
where $J_n$ is the Bessel function of the first kind~\cite{Hartmann_2004}. Introducing a hard-wall edge, the exact single-particle wave function can be obtained as
\begin{align}
    \psi_n(t) = i^{\,n-1}\left[ J_{n-1}(2Jt) + J_{n+1}(2Jt) \right]
              = i^{\,n-1} \left(\frac{n}{Jt}\right) J_n(2Jt).
    \label{bessel}
\end{align}

In our simulation, the doublon and the holon propagate on the half-filled background as free particles with hopping $J$, so that the wave packets before the collision are described by Eq.~\eqref{bessel}.
\begin{figure*}[!b]
\centering
\includegraphics[width=\textwidth]{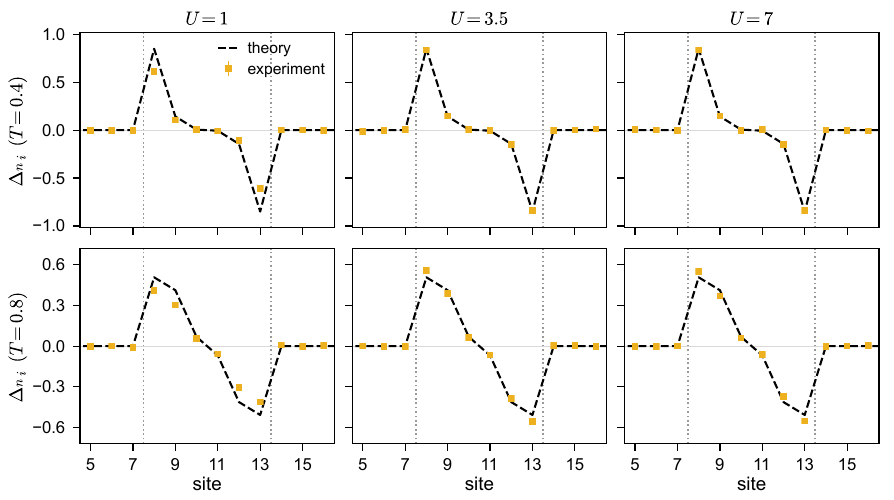}
\caption{Comparison of the experimental site-resolved charge density in Phase~1 with the theoretical Bessel-function solution of the semi-infinite chain. Yellow markers are the experimental values, black dashed lines are the theoretical values, and gray vertical dashed lines mark the positions of the deactivated hoppings.}
\label{figs1}
\end{figure*}

\begin{figure*}[t]
\centering
\includegraphics[width=\textwidth]{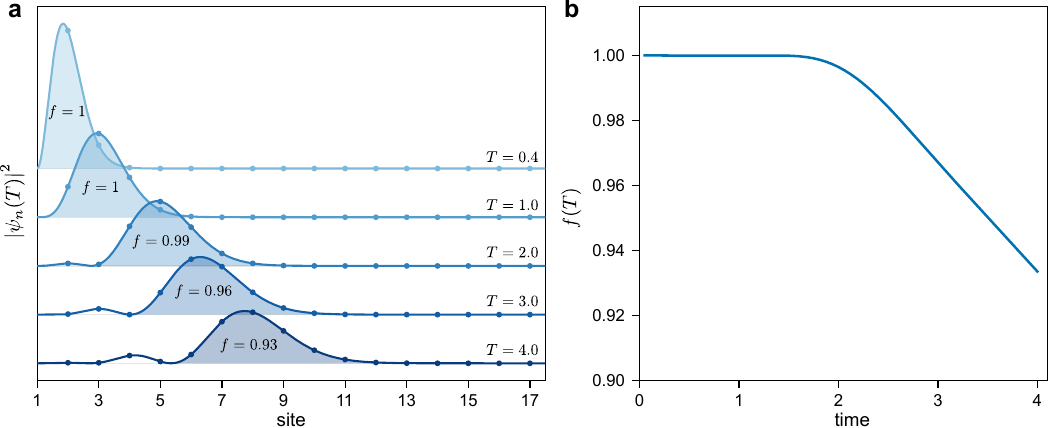}
\caption{Spatial profile of the semi-infinite chain solution as a function of time (left) and the contribution of the dominant packet as a function of time (right). Profiles are shown for $T=0.4,\,1.0,\,2.0,\,3.0,\,4.0$, and $f$ is defined as the fraction of the total wave function occupied by the first peak (integral of the shaded region). The right panel shows the time dependence of $f$.}
\label{figs2}
\end{figure*}

Figure~\ref{figs1} shows the directed quasiparticle wave packets of the charge density in our experimental setup (yellow markers), compared to the wave function of Eq.~\eqref{bessel} (black dashed lines) during phase 1. The packet shapes agree better at large $U$, where the quasiparticles are well defined, while charge fluctuations distort the packets at small $U$. The wave-function profile of Eq.~\eqref{bessel} consists of a dominant leading packet followed by weaker trailing packets. Within our simulation window, $T \le 4$, we assume that the leading packet is sufficiently dominant for the trailing packets to be ignored. Figure~\ref{figs2} shows the extent to which the leading packet dominates. Figure~\ref{figs2}(a) displays the wave-function profile $|\psi_n(T)|^2$ of Eq.~\eqref{bessel}, with the contribution of the dominant leading packet shaded. Figure~\ref{figs2}(b) plots the leading-packet contribution $f(T)$. For all simulation times, more than 93\% of the weight is concentrated in the leading packet, and the contributions of the trailing packets are therefore negligible on our simulation time scale.

\section{Qubit mapping and circuit implementation}\label{sec:circuit}

In this section, we describe the qubit mapping and the circuit construction in detail. The spin-up and spin-down orbitals of each site $s$ are assigned to the qubit pair $(2s-1,\,2s)$. We adopt a zigzag mapping in which the order within each pair is $(\uparrow,\downarrow)$ on even sites and $(\downarrow,\uparrow)$ on odd sites, so that orbitals carrying the same spin species on adjacent sites are mapped onto neighboring qubits and the hopping terms require no long-range gates~\cite{hartnett2026fastaccuratehighresolutionsimulation}. Arranging the fermionic modes in this zigzag order, we apply the Jordan--Wigner transformation~\cite{JWtransform}
\begin{align}
    c_j = \Big(\prod_{k<j} Z_k\Big)\,\sigma^-_j .
\end{align}
In this ordering, every term of the Hamiltonian connects only adjacent modes, between which the Jordan--Wigner strings cancel; the hopping terms
therefore reduce to the local two-qubit operations
\begin{align}
    c^\dagger_q c_{q+1} + \mathrm{h.c.}
    \;\longrightarrow\;
    \tfrac{1}{2}\big(X_q X_{q+1} + Y_q Y_{q+1}\big).
\end{align}
The even bonds $(2s,\,2s+1)$ connect the two spin orbitals of the same site and carry the onsite interaction together with the fermionic swap~\cite{PhysRevLett.120.110501}, while the odd bonds $(2s+1,\,2s+2)$ implement the hopping between neighboring sites. In a given spin convention, a single hopping layer implements half of the hopping terms and the remaining half is implemented with an identical gate arrangement by the next hopping layer, after the fermionic swaps exchange the convention [see Fig.~\ref{fig:setup}(e)].

The time evolution is implemented by a modified second-order Trotter decomposition of the three-term splitting of the hopping and interaction
parts, with a single step defined as
\begin{align}
    U_{\rm step}(dt) \;=\; K(dt)\; W(dt)\; K(dt).
\end{align}
Here, $K(dt)$ applies the hopping gate
\begin{align}
    U_{\rm hop}(J,dt) = \exp\!\left[i\,\tfrac{Jdt}{2}\big(XX+YY\big)\right],
\end{align}
to every odd bond; owing to the fermionic swaps contained in $W$, the two hopping layers of each step address the two different spin species of every site. $W(dt)$ applies to every even bond (all $L$ site pairs) $U_\text{int}$ which is the gate combining the fermionic swap with the on-site interaction,
\begin{align}
    U_\text{int}(U,dt) \;=\; F \cdot \mathrm{diag}\big(1,\ 1,\ 1,\ e^{-i U dt}\big),
    \qquad
    F = \begin{pmatrix} 1 & 0 & 0 & 0 \\ 0 & 0 & 1 & 0 \\ 0 & 1 & 0 & 0 \\ 0 & 0 & 0 & -1 \end{pmatrix},
\end{align}
where $F$ denotes the fermionic swap. In consecutive steps, the last $K(dt)$ layer and the first $K(dt)$ layer of the next act on the same set of hopping terms and are merged into a single layer $K(2dt)$ with time step $2dt$. The hopping gate $U_\text{hop}$is decomposed into two CZ gates with single-qubit rotations, 

\begin{align}
U_{\rm hop} \;=\;
\begin{quantikz}[column sep=0.12cm, row sep=0.35cm]
& \gate{R_x\!\big(\tfrac{\pi}{2}\big)} & \gate{H} & \ctrl{1} & \gate{H} & \gate{R_z(-J\,dt)} & \gate{H} & \ctrl{1} & \gate{H} & \gate{R_x\!\big(-\tfrac{\pi}{2}\big)} & \\
& \gate{R_x\!\big(\tfrac{\pi}{2}\big)} &          & \control{} &        & \gate{R_x(-J\,dt)} &          & \control{} &        & \gate{R_x\!\big(-\tfrac{\pi}{2}\big)} &
\end{quantikz}
\;=\;
\begin{pmatrix}
1 & 0 & 0 & 0 \\
0 & \cos(J\,dt) & i\sin(J\,dt) & 0 \\
0 & i\sin(J\,dt) & \cos(J\,dt) & 0 \\
0 & 0 & 0 & 1
\end{pmatrix}
\label{eq:uhop_circuit}
\end{align}

and $U_\text{int}(U,dt)$ into three CZ gates with single-qubit rotations up to a global phase.

\begin{align}
U_{\rm int} \;=\;
\begin{quantikz}[column sep=0.12cm, row sep=0.35cm]
& \gate{H} & \ctrl{1} & \gate{H} & \ctrl{1} & \gate{R_z\!\big(\tfrac{U dt}{2}-\tfrac{\pi}{2}\big)} & \gate{H} & \ctrl{1} & \gate{H} & \gate{R_z\!\big(\tfrac{\pi}{2}-\tfrac{U dt}{2}\big)} & \\
&          & \control{} & \gate{H} & \control{} & \gate{H} &        & \control{} &        & \gate{R_z\!\big(\tfrac{\pi}{2}-\tfrac{U dt}{2}\big)} &
\end{quantikz}
\;=\;
e^{\,i\frac{U dt-\pi}{4}}
\begin{pmatrix}
1 & 0 & 0 & 0 \\
0 & 0 & 1 & 0 \\
0 & 1 & 0 & 0 \\
0 & 0 & 0 & -e^{-iU dt}
\end{pmatrix}
\label{eq:uint_circuit}
\end{align}

The $N$-step Trotter circuit consists of $N+1$ hopping layers and $N$ interaction layers, amounting to $2(N+1) + 3N = 5N + 2$ two-qubit gate layers; our experiments run up to $N = 20$, corresponding to a two-qubit depth of $102$.

\section{Error suppression and mitigation}

This section describes the details and the validation of the error suppression and error mitigation methods used in the hardware experiment.

\subsection{Qubit layout selection}
\begin{figure}[t]
\centering
\includegraphics[width=\textwidth]{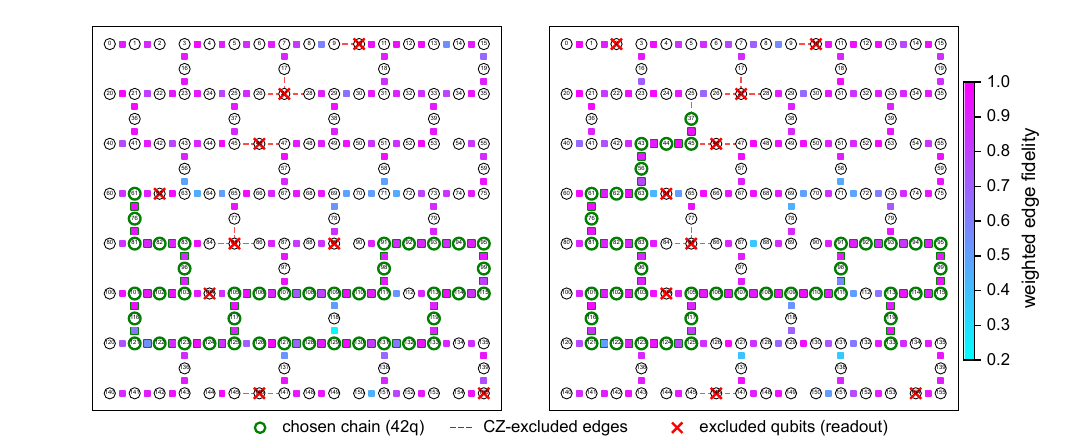}
\caption{Selection of the optimal layout used in the two experimental sessions. Each circle denotes an available qubit of ibm\_boston, and the color of the square on each edge denotes the fidelity obtained from the Bell test. Qubits and edges highlighted in green were selected for the experiment; qubits marked with a red X were excluded before the test due to readout errors, red dashed lines are edges excluded before the test due to CZ errors, and empty edges are those excluded due to a mismatch in two-qubit gate duration.}
\label{figS3}
\end{figure}

Placing the 42-qubit linear chain corresponding to $L=21$ sites optimally on the 156-qubit heavy-hex lattice is essential to minimize the influence of hardware noise. In particular, in a one-dimensional chain a single faulty coupler corrupts the entire hopping term at that position, so it is important that not even one very bad qubit or coupler is included in the chain. To this end we performed a three-stage selection within every experimental session.

Using the backend calibration data provided before benchmarking, we excluded from the candidates the qubits with readout error above 0.02, the couplers with calibrated CZ error above 0.01, and the couplers whose two-qubit gate duration differs from the majority of couplers (68~ns). We then performed a Bell state test~\cite{mckay2023benchmarkingquantumprocessorperformance, cosentino2026metriqcollaborativeplatformbenchmarking} on all remaining couplers by applying the Hardarmard gate and CX gate 15 times repeatedly and measuring with 2048 shots, and used for the experiment the path with the largest product of the per-edge fidelities between the measured and the ideal distributions. Fig.~\ref{figS3} shows the coupler fidelity maps measured in the two sessions and the selected chains.

\subsection{Trotter-error mitigation}

In quantum simulation one has to consider not only the hardware error but also the Trotter error, or algorithmic error, that arises from splitting the time evolution into finite time steps. When both errors are considered together for a given time evolution, a shorter time step $\Delta t$ reduces the algorithmic error but increases the influence of the hardware noise, so it is important to analyze the trade-off between the two and find the optimal time step~\cite{bark2026constantdepthmultiproductformulatrotter}.

\begin{figure}[t]
\centering
\includegraphics[width=\textwidth]{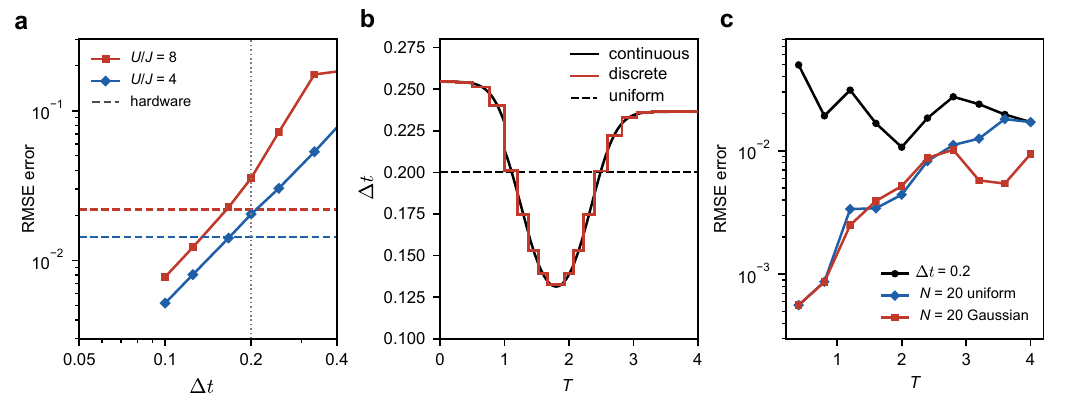}
\caption{Trotter error and setting of the optimal time step. (a) Trotter error (solid line, MPS calculation) and hardware noise at a two-qubit depth of 102 (dashed line) for $T=4$. The error is the RMSE of the charge density averaged over sites. The gray vertical dashed line marks the time step corresponding to the number of steps adopted in the experiment ($N=20$). (b) Time steps used in the $T=4$ experiment. The black solid line is the continuous density function $\rho(t)$, the red solid line is the list of time steps actually applied in the experiment, obtained by discretizing into 20 steps, and the black dashed line is the uniform time step. (c) Comparison of the Trotter error of the three schedules as a function of evolution time at $U/J=7$. Black: $\Delta t=0.2$ fixed with a varying number of steps; blue: uniform time steps with $N=20$ for every time; red: the Gaussian time steps used in the experiment. The error is the RMSE of the charge density.}
\label{figS4}
\end{figure}

Fig.~\ref{figS4}(a) shows the Trotter error of the observables at $T=4$ when a uniform Trotter step is used. The errors are obtained by running the 21-site circuit in an MPS simulation as a function of the uniform time step $\Delta t=T/N$, and show the expected gradual increase as $\Delta t$ becomes larger. The horizontal dashed lines show the hardware error measured from a circuit of the same depth, 102 layers. This hardware error is computed as the RMSE between a preliminary experimental run with uniform $\Delta t=0.2$ and the corresponding noiseless simulation under the same conditions, therefore, it reflects only the hardware noise and excludes the Trotter error.
The hardware error provides a practical reference scale: once the Trotter error exceeds this level, it becomes the dominant contribution to the total error. Thus, the most effective strategy is to choose a Trotter step for which the Trotter error is smaller than, or at least comparable to, the hardware error. Making the Trotter error much smaller would require unnecessary circuit depth, while allowing it to become larger would make the Trotter error the dominant cost. Based on this balance, we fixed $N=20$ steps, equivalent to the time step $\Delta t=0.2$, where the hardware error and Trotter error are comparable, for all simulations.

Since we fixed the number of steps rather than the time step, we used $N=20$ also for short measurement times. This is to remove the influence of the Trotter error, which is hard to predict, as much as possible while accepting a sufficiently tolerable level of hardware noise. Within each circuit we then redistributed the step sizes in time instead of keeping them uniform. We defined the step density function
\begin{align}
\rho(t) \propto 1 + A\exp\!\left[-\frac{(t-t_c)^2}{2\sigma^2}\right],
\label{eq:step_density}
\end{align}
discretized the optimized density function into 20 intervals while keeping the number of steps and the total evolution time fixed, and ran the simulation with a time-dependent $\Delta t$. The resulting time steps are small near the collision center $t_c$ and larger elsewhere, which is motivated by the physical properties of the quasiparticle dynamics. The motion of a single doublon or holon in the half-filled N\'eel background is a first-order hopping to the neighboring site, so its front velocity is theoretically $2J$ and hardly depends on $U$ except through the channels in which the quasiparticle decays. In contrast, at the moment the doublon and the holon overlap, the annihilation, transmission, and reflection channels branch according to $U/J$ and the $U$ dependence returns. Since the effect of the $U$-dependent quasiparticle decay can be partially canceled by evolving the background together and subtracting it, if only the Trotter error due to large $U$ near the collision time $t=t_c$ is considered, making the steps denser around the collision than in the other parts of the dynamics can be effective in reducing the Trotter error.

The parameters used in our simulations are $A=0.8$, $\sigma=0.4$, and $t_c=1.8$, and a single schedule was applied to all $U$ in common. Fig.~\ref{figS4}(b) shows the continuous function $\Delta t(t)$ at $T=4$, the 20 steps actually executed, and the fixed step $\Delta t=0.2$ for comparison. The schedule was optimized from the MPS simulation of the 21-site circuit at $U/J=8$ and $T=4$, and all other simulation results were obtained with the same schedule. Fig.~\ref{figS4}(c) compares, at $U/J=7$, the Trotter errors of three schedules as a function of the measurement time: the fixed step $\Delta t=0.2$, uniform $N=20$, and $N=20$ optimized with the Gaussian form. One can see that the Trotter error is low at short times because many steps are used there instead of fixing the time step, and that the Gaussian redistribution reduces the Trotter error significantly in the long-time region.

\subsection{Pauli twirling, dynamical decoupling, and readout twirling}

Pauli twirling inserts random Pauli operators before and after each two-qubit gate, paired so that the two operators cancel logically, and converts the coherent error of the gate into a stochastic Pauli channel in the average over many instances ~\cite{PhysRevA.94.052325}.
In our experiment each circuit was compiled into 70 random Pauli instances with the \texttt{Samplomatic} software library. We also applied dynamical decoupling (XY4), which inserts an XYXY pulse sequence into the idle periods without gates, to cancel low-frequency phase noise ~\cite{PhysRevA.58.2733,PhysRevLett.121.220502,PhysRevApplied.20.064027}.
Finally, immediately before the measurement we applied readout twirling, which inserts random X gates and undoes the corresponding flips in post-processing. This converts the state-dependent and asymmetric readout error into a symmetric bit-flip channel on each qubit, and was performed in order to apply twirled readout error extinction (TREX)~\cite{PhysRevA.105.032620,doi:10.1126/sciadv.abi8009}.
To estimate the readout-error-dependent attenuation factors for TREX, we also measured a calibration circuit with all qubits in $\ket{0}$ under the same readout twirling and computed the per-qubit attenuation $\gamma_q=\braket{Z_q}_{\mathrm{cal}}$. However, since the subsequent mitigation reads the total flip probability including the readout error and corrects for it together, adding the extinction step of dividing by $\gamma_q$ changed the mitigated observables only marginally.

\subsection{Symmetry-based restoration of expectation values}

\begin{figure}[t]
\centering
\includegraphics[width=\textwidth]{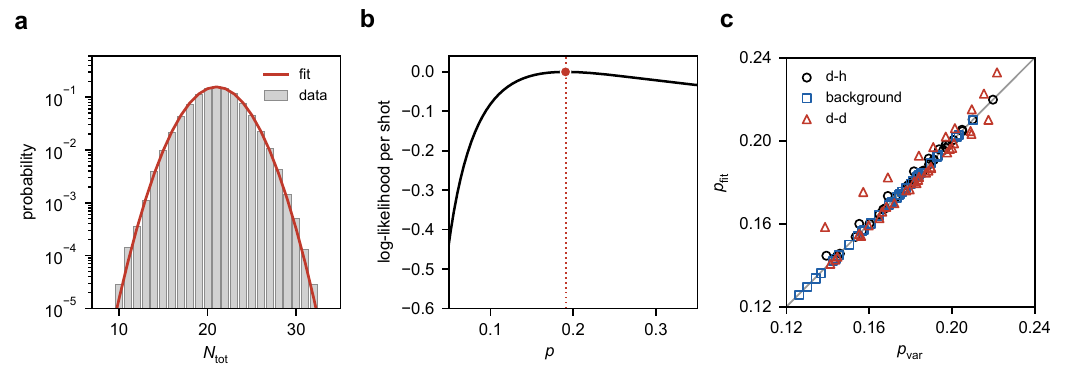}
\caption{Estimation of the circuit noise parameter from the particle-number distribution. (a) Histogram of the total particle number $N_{\mathrm{tot}}$ of each shot (gray, log scale) for 70,000 shots of the experiment at $U/J=3.5$, $T=3.2$, and the distribution at the value of $p$ fitted with the uniform error model (red solid line). (b) $p$ dependence of the per-shot log-likelihood in the uniform-error-model fit. The red dashed line and marker indicate the optimal $p$. (c) Comparison of $p_{\mathrm{fit}}$ obtained from the fit and $p_{\mathrm{var}}$ obtained from the variance formula for all circuit results used in the experiment (111 in total). The closer the two values (closer to $y=x$), the better the experimental result is approximated by the simple uniform error model.}
\label{figS5}
\end{figure}

Once a sufficiently good layout is chosen and arbitrary noise is converted into a Pauli channel by Pauli twirling, local errors spread sufficiently in a sufficiently deep circuit, so that for the $Z$-basis expectation values the noise can be approximated by a flip error with probability $p$ distributed uniformly over all qubits~\cite{noise_eff, PhysRevE.104.035309,PhysRevLett.127.270502}. Noting that the simulation circuit exactly conserves the total particle number $N_{\mathrm{tot}}$ and that the total particle number of the initial state is known, it can give useful information which is regarded on the hardware noise~\cite{PhysRevA.98.062339, Cai_2021}. We can estimate the bit-flip probability from the distribution of the total particle number. In the uniform bit-flip approximation, a flip of one of the $N_0$ occupied qubits loses a particle and a flip of one of the $N_Q-N_0$ empty qubits gains a particle, so the measured total particle number follows the distribution
\begin{align}
N' = N_0 - \mathrm{Bin}(N_0,p) + \mathrm{Bin}(N_Q-N_0,p), \qquad N_Q=42.
\label{eq:ntot_dist}
\end{align}
We determined $p$ by fitting this distribution to the histogram of the total particle number obtained from the 70,000 shots of each data point (Fig.~\ref{figS5}(a), (b)). In the ideal situation the variance of the distribution is $\mathrm{Var}(N')=N_Q\,p(1-p)$, so $p$ is also determined in closed form from the variance alone. Comparing $p_{\mathrm{fit}}$ obtained from the fit with $p_{\mathrm{var}}$ obtained from the variance therefore allows us to check whether the approximation we use is adequate. Comparing the two values for all 111 $(U,T)$ combinations (Fig.~\ref{figS5}(c)), the two estimates lie mostly on the line $y=x$, which confirms that the approximation we use describes the actual experimental data well. With the global bit-flip probability $p$ obtained in this way, we can estimate the ideal expectation values inversely from the measured noisy expectation values. When the error channel has bit-flip probability $p$, the noisy expectation value $\braket{b_{i,\sigma}}_{\mathrm{noisy}}$ of the particle occupation of the qubit representing site $i$ and spin $\sigma$ is related to the ideal expected occupation $\braket{b_{i,\sigma}}_{\mathrm{ideal}}$ as
\begin{align}
\braket{b_{i,\sigma}}_{\mathrm{noisy}} = (1-2p)\,\braket{b_{i,\sigma}}_{\mathrm{ideal}} + p.
\label{eq:b_noisy}
\end{align}
Defining the coefficient $\lambda=(1-2p)$, the expectation values of the charge $n_i=b_{i,\uparrow}+b_{i,\downarrow}$, the doublon $d_i=b_{i,\uparrow}b_{i,\downarrow}$, and the holon $h_i=(1-b_{i,\uparrow})(1-b_{i,\downarrow})$ of site $i$, defined through its two qubit occupations $b_{i,\uparrow}$ and $b_{i,\downarrow}$, are expressed in terms of $p$ as
\begin{align}
\braket{n}_{\mathrm{noisy}} &= \lambda\braket{n}_{\mathrm{ideal}} + 2p, \\
\braket{d}_{\mathrm{noisy}} &= \lambda^2\braket{d}_{\mathrm{ideal}} + \lambda p\,\braket{n}_{\mathrm{ideal}} + p^2, \\
\braket{h}_{\mathrm{noisy}} &= \lambda^2\braket{h}_{\mathrm{ideal}} - \lambda p\,\braket{n}_{\mathrm{ideal}} + (2p-3p^2).
\label{eq:ndh_noisy}
\end{align}
Since we know the noisy values obtained from the experiment and $p$, or $\lambda$, obtained from the particle-number distribution of the bitstrings, the ideal values can be restored as
\begin{align}
\braket{n}_{\mathrm{ideal}} &= \frac{\braket{n}_{\mathrm{noisy}} - 2p}{\lambda}, \\
\braket{d}_{\mathrm{ideal}} &= \frac{\braket{d}_{\mathrm{noisy}} - p\,\braket{n}_{\mathrm{noisy}} + p^2}{\lambda^2}, \\
\braket{h}_{\mathrm{ideal}} &= \frac{\braket{h}_{\mathrm{noisy}} + p\,\braket{n}_{\mathrm{noisy}} - 2p + p^2}{\lambda^2}.
\label{eq:ndh_restore}
\end{align}
In the main text, we primarily examine the results with quasiparticle excitation after subtracting the corresponding results without quasiparticle excitation, in order to remove the influence of the background. If the two cases have the same circuit error rate, the constant terms cancel and the result can be expressed simply as
\begin{align}
\Delta\braket{n}_{\mathrm{ideal}} &= \frac{\Delta\braket{n}_{\mathrm{noisy}}}{\lambda}, \\
\Delta\braket{d}_{\mathrm{ideal}} &= \frac{\Delta\braket{d}_{\mathrm{noisy}} - p\,\Delta\braket{n}_{\mathrm{noisy}}}{\lambda^2}, \\
\Delta\braket{h}_{\mathrm{ideal}} &= \frac{\Delta\braket{h}_{\mathrm{noisy}} + p\,\Delta\braket{n}_{\mathrm{noisy}}}{\lambda^2}.
\label{eq:delta_restore}
\end{align}
In the actual experiment, however, the probability may differ between the circuits, so we take the approach of restoring each observable first and then taking the difference. Since $\braket{n}=1$ is a fixed point that is unchanged by this error channel and by the restoration, a large improvement may not occur for the background alone or in the intermediate-interaction region where $\Delta n$ is small and flat. The mitigation performance is discussed together in the next section.

\subsection{Symmetry-based cancellation of fluctuations}

\begin{figure}[t]
\centering
\includegraphics[width=\textwidth]{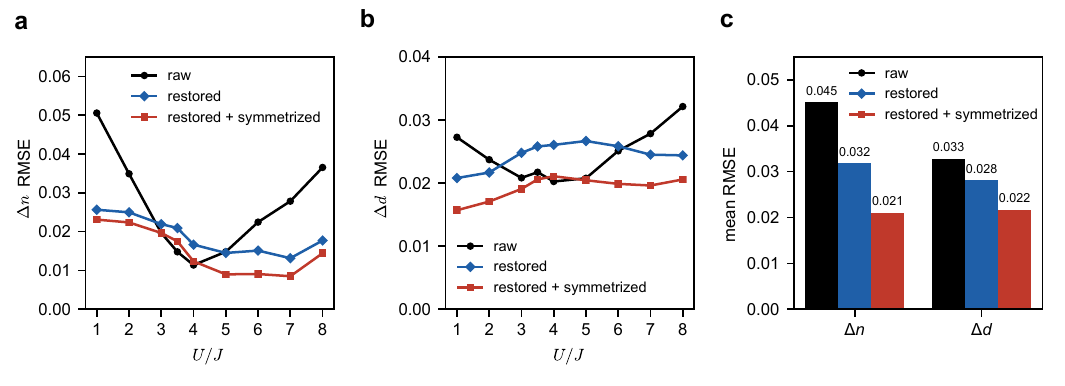}
\caption{Error mitigation performance of the symmetry-based restoration and symmetrization. The converged MPS result is used as the reference, and the error is the site RMSE of each operator. (a, b) $U$ dependence of the error of the charge density $\Delta n$ (a) and the doublon density $\Delta d$ (b) at $T=3.2$. Black lines are the raw data, blue lines are after restoration, and red lines are after applying both symmetrization and restoration. (c) Bar plot of the site RMSE averaged over all $(U/J,\,T)$ combinations used in the experiment for each error mitigation method.}
\label{figS6}
\end{figure}

The method above focused on restoring the observables reduced by the errors, but the approximation may not hold well when the error is nonuniform across the sites, or when there is an asymmetric error in which 0 and 1 flip with different probabilities. To mitigate this effect we use another symmetry that the Hamiltonian and our initial state possess. Under uniform hopping and half filling, the Hamiltonian and the initial state are invariant under the composition $\mathcal{I}\mathcal{C}$ of the inversion about the chain center, $\mathcal{I}: i\mapsto L+1-i$, and the particle-hole transformation $\mathcal{C}: c_{i\sigma}\mapsto(-1)^i c^\dagger_{i\sigma}$. Therefore, at any time the ideal observables satisfy
\begin{align}
\Delta n_i = -\Delta n_{L+1-i}, \qquad \Delta d_i = \Delta h_{L+1-i}.
\label{eq:phmirror_relations}
\end{align}
We can therefore define the symmetrized observables as
\begin{align}
\Delta n_i &\rightarrow \frac12\bigl(\Delta n_i - \Delta n_{L+1-i}\bigr), \\
\Delta d_i &\rightarrow \frac12\bigl(\Delta d_i + \Delta h_{L+1-i}\bigr), \\
\Delta h_i &\rightarrow \frac12\bigl(\Delta h_i + \Delta d_{L+1-i}\bigr).
\label{eq:symmetrized}
\end{align}
In our experiment the last site, which has no partner, was excluded from the symmetrization, and we confirmed with tensor network calculations that the effect of this exclusion is negligible. The doublon--doublon initial state does not satisfy the particle-hole symmetry, but it satisfies the inversion symmetry with global spin flip, so the symmetrization can be carried out in the same way. Rather than reducing the noise-induced error by itself, the symmetrization has the effect of reducing the variance of the components that fluctuate independently from site to site and of the state-dependent nonuniform flip rates. In this experiment the symmetrization therefore complements the weakness of the restoration method, which can restore the reduced component of the observables but is vulnerable to nonuniform fluctuations. Fig.~\ref{figS6} compares the error rates for the symmetry-based error mitigation methods. The tensor network results were used as the reference values. The figure shows that the symmetry-based error mitigation improves the error rate in most of the parameter range. As mentioned above, restoration alone can amplify the error in regions where $\Delta n$ is small, whereas symmetrization stabilizes the variance and improves the mitigation quality.

\section{Tensor network benchmarking}

\begin{figure}[t]
\centering
\includegraphics[width=\textwidth]{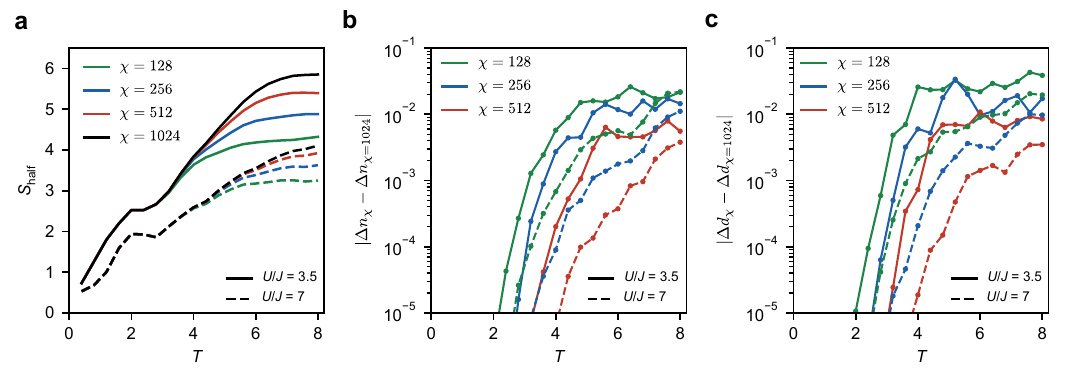}
\caption{Convergence check of the tensor network calculation as a function of evolution time. (a) Time dependence of the entanglement entropy at the central bond of the chain for $\chi=128$ (green), $256$ (blue), $512$ (red), and $1024$ (black); solid and dashed lines correspond to $U/J=3.5$ and $7$, respectively. (b, c) Error of the charge (b) and doublon (c) density of the lower-$\chi$ results relative to the $\chi=1024$ result. Within the range measured in the experiment ($T\leq 4$), the $\chi$ dependence remains below $10^{-3}$ for $\chi\geq 512$.}
\label{figS7}
\end{figure}

We performed tensor-network-based TEBD simulations to validate the effect of the error mitigation, the Trotter error, and the experimental results. The code was implemented with the simulation toolkit \texttt{TeNPy} ~\cite{tenpy2024},
and the maximum bond dimension was kept within $\chi=1024$. In the range $T\le4$ measured in the experiment, the results for $\chi=256$, $512$, and $1024$ agree within $10^{-3}$ for both the site density and the doublon density, confirming convergence. To quantify the time at which convergence becomes difficult, we repeated the same calculation up to $T=8$ and tracked the entanglement entropy $S_{\mathrm{half}}$ and the $\chi$ dependence of each observable (Fig.~\ref{figS7}). Beyond $T=4$ the entanglement entropy did not converge, and the maximum deviation of the site density $\Delta n$ between $\chi=512$ and $1024$ stayed below $10^{-3}$ for $T\le4$ and grew gradually afterwards.

\section{Bootstrapping for error bars}

All error bars were obtained by bootstrapping. We constructed a multinomial distribution from the experimentally obtained bitstrings, resampled 70,000 shots with replacement independently for each resample, recomputed the post-processing (mitigation) from scratch for each resample to obtain 1000 replicates, and took the standard deviation as the error bar.

\section{Theory of doublon--holon scattering}
 
In addition to the spin SU(2), the Fermi--Hubbard Hamiltonian commutes with the $\eta$-pairing SU(2) generated by $\eta^{+}=\sum_{i}(-1)^{i}c^{\dagger}_{i\uparrow}c^{\dagger}_{i\downarrow}$ and $\eta^{z}=\tfrac12\sum_{i}(n_{i\uparrow}+n_{i\downarrow}-1)$~\cite{PhysRevLett.63.2144}.
The half-filled ground state is an SO(4) singlet, and the charge excitations above it, the holon and the doublon or antiholon form a pseudospin doublet~\cite{PhysRevLett.72.908}. The scattering of a doublon--holon pair therefore has the same structure as the scattering of two spin-$\tfrac12$ particles, and its exact solution is obtained by the Bethe ansatz~\cite{7ef6589804f8480e8e84711e5ad18a16}.

For the state $\ket{\Psi}=\sum_{x_{1},x_{2}}\psi(x_{1},x_{2})\, c^{\dagger}_{x_{1}\uparrow}c^{\dagger}_{x_{2}\downarrow}\ket{0}$, the Schr\"odinger equation can be written as
\begin{align}
E\,\psi(x_{1},x_{2})=-J\big[\psi(x_{1}{+}1,x_{2})+\psi(x_{1}{-}1,x_{2})+\psi(x_{1},x_{2}{+}1)+\psi(x_{1},x_{2}{-}1)\big]+U\,\delta_{x_{1},x_{2}}\,\psi(x_{1},x_{2}).\label{eq:S_schroedinger}
\end{align} 
The Bethe ansatz expresses the wave function as a superposition of two plane waves,
\begin{align}
\psi(x_{1},x_{2})&=A_{12}\,e^{i(k_{1}x_{1}+k_{2}x_{2})}+A_{21}\,e^{i(k_{2}x_{1}+k_{1}x_{2})}\quad(x_{1}<x_{2}),\nonumber\\ \psi(x_{1},x_{2})&=B_{12}\,e^{i(k_{1}x_{1}+k_{2}x_{2})}+B_{21}\,e^{i(k_{2}x_{1}+k_{1}x_{2})}\quad(x_{1}>x_{2}).
\label{eq:S_bethe_ansatz}
\end{align}
Since the interaction acts only on the $x_{1}=x_{2}$, Eq.~\eqref{eq:S_schroedinger} reduces to the free equation inside each region, and every plane wave satisfies $E=-2J(\cos k_{1}+\cos k_{2})$. Imposing continuity $A_{12}+A_{21}=B_{12}+B_{21}$ together with the matching condition
\begin{align}
-2iJ\left(\sin k_{1}-\sin k_{2}\right)\left(B_{12}-A_{12}\right)+U\left(A_{12}+A_{21}\right)=0,
\label{eq:S_matching}
\end{align}
choosing $\sin k_{1}>\sin k_{2}$ and $B_{21}=0$, the scattering coefficients can be obtained as
\begin{align}
  t\equiv\frac{B_{12}}{A_{12}}=\frac{\theta}{\theta+i},\qquad
  r\equiv\frac{A_{21}}{A_{12}}=\frac{-i}{\theta+i},\qquad
  \theta\equiv\frac{2J\left(\sin k_{1}-\sin k_{2}\right)}{U} .
  \label{eq:S_tr}
\end{align}
The $S$-matrix~\cite{PhysRevLett.19.1312} is therefore
\begin{align}
  S(\theta)=\frac{\theta\ -i\ \Pi}{\theta+i},
  \label{eq:S_Smatrix}
\end{align}
where $\Pi$ is the exchange operator that swaps the momenta of the two particles. Although this is a result for the empty chain, the exact charge-sector $S$-matrix of the half-filled Fermi--Hubbard model ~\cite{PhysRevLett.72.908} has the same form as Eq.~\eqref{eq:S_Smatrix} for the same $\theta$ up to an overall phase due to the Mott background. Hence the transmission and reflection probabilities for a doublon--holon pair at half filling are given by
\begin{align}
  \mathcal{T}=|t|^{2}=\frac{\theta^{2}}{1+\theta^{2}},\qquad
  \mathcal{R}=|r|^{2}=\frac{1}{1+\theta^{2}}.
\end{align}

\section{Entanglement probe based on the particle-number fluctuation} \label{entangle}

We divide the chain at its center into the left half $L$ and the right half $R$. We assume that the total state $\ket{\psi}$ is pure and consider the case in which the circuit exactly conserves the total particle number $\hat N=\hat N_L+\hat N_R$. In this case the reduced density matrix can be block-diagonalized with respect to the particle number $N_L$ of the left half of the chain as
\begin{align}
\rho_L=\bigoplus_{n}P(n)\,\rho_L^{(n)}.
\label{eq:rhoL_blocks}
\end{align}
Here $\rho_L$ is the reduced density matrix of the left half of the chain, and $P(n)$ is the probability of finding $n$ particles when the left half is measured, which can be obtained as the trace of the block with $N_L=n$. 

The entanglement entropy $S(\rho_L)$ can be decomposed into the contribution $S_{\mathrm{num}}$ from the particle-number distribution and the contribution $S_{\mathrm{block}}^{(n)}$ from the entanglement structure inside each block-diagonal matrix, as
\begin{align}
S(\rho_L)=S_{\mathrm{num}}+\sum_n P(n)\,S_{\mathrm{block}}^{(n)}
=-\sum_n P(n)\ln P(n)+\sum_n P(n)\,S\!\left(\rho_L^{(n)}\right).
\label{eq:S_decomposition}
\end{align}
Since both terms are non-negative, the Shannon entropy $S_{\mathrm{num}}$ of the particle-number distribution is a lower bound on the total entanglement entropy ~\cite{PhysRevLett.102.100502,PhysRevB.82.012405,PhysRevB.85.035409,doi:10.1126/science.aau0818}. In the actual experiment, however, the particle number can change due to noise, and when the final state is a mixed state the lower bound cannot be established in this way. In this work, we therefore use it as a tool for observing qualitative features rather than a rigorous lower bound on the entanglement entropy. 

The quantity we measure is the variance $\mathrm{Var}(N_L)$ of the particle-number distribution of the left half~\cite{PhysRevB.85.035409}. This quantity is not exactly the same as the Shannon entropy, but if we assume that the state after the collision consists of two components, a transmitted and a reflected state, the particle number of the left half differs by 2 between them, so that for the transmission probability $\mathcal{T}$ the Shannon entropy and the variance can be written as
\begin{align}
S_{\mathrm{num}}=-\mathcal{T}\ln\mathcal{T}-(1-\mathcal{T})\ln(1-\mathcal{T}),\qquad
\Delta\mathrm{Var}(N_L)=4\mathcal{T}(1-\mathcal{T}).
\label{eq:two_component}
\end{align}
Both quantities are monotonically increasing functions of $\mathcal{T}(1-\mathcal{T})$, so they correspond monotonically to each other, and the $U$ dependence of the variance can reflects the $U$ dependence of $S_{\mathrm{num}}$, which provides the lower bound~\cite{PhysRevLett.102.100502,PhysRevB.82.012405,PhysRevB.85.035409}. In contrast, for the doublon--doublon collision $S_{\mathrm{num}}$ stays close to zero even after the collision because of the exchange symmetry of identical particles. The half-filled N\'eel background is not an eigenstate, so its particle-number fluctuation can grow in time. Since independent contributions simply add in the variance, however, in this work we compute the variances of the collision and of the background separately and compare their difference.

\begin{figure}[t]
\centering
\includegraphics[width=\textwidth]{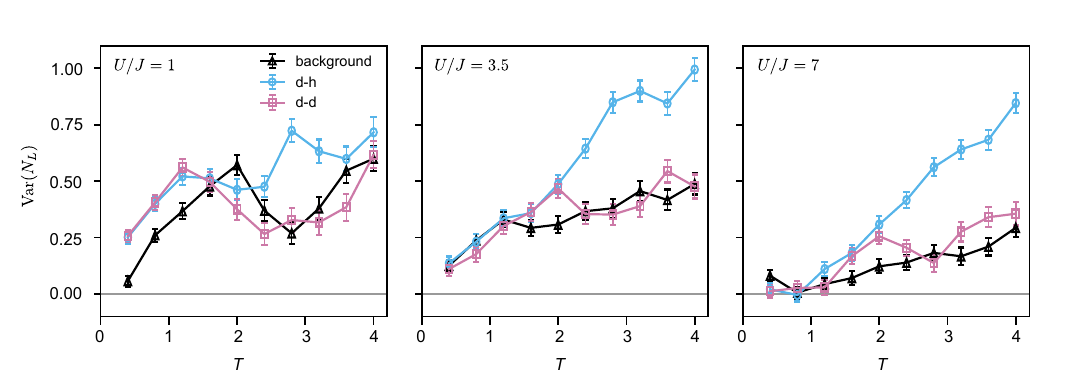} 
\caption{Particle-number fluctuation $\mathrm{Var}(N_L)$ of the left half for three interaction strengths ($U/J=1$, $U/J = 3.5$ and $U/J=7$) as a function of time. The black, blue, and pink plots denote $\mathrm{Var}(N_L)$ for the N\'eel background, the doublon--holon(d-h) collision, and the doublon--doublon(d-d) collision, respectively.}
\label{figS8}
\end{figure}

Fig.~\ref{figS8} shows the time dependence of $\mathrm{Var}(N_L)$ for the background, the doublon--holon(d-h) collision, and the doublon--doublon(d-d) collision. At $U/J=3.5$, where the entanglement is strong, all three values grow together before the collision ($T\lesssim2.0$), and after the collision the d-h case increases steeply, unlike the d-d case, which remains almost identical to the background. At $U/J=7$ the background fluctuation is small, so the variance grows more slowly than for the other two interaction strengths, and the d-h case has the larger variance after the collision. Finally, at $U/J=1$ the quasiparticles are not well defined, and the effect of the background fluctuation dominates over the effect of the collision, so the collision effect is not observed as clearly as for the other interaction strengths.

